\documentclass[10pt]{article}
\usepackage[OE]{express}
\usepackage{color,soul}
\begin{document}

\title{Large-scale spectral bandwidth compression by complex electro-optic temporal phase modulation}

\author{Filip So{\'s}nicki\authormark{1} and Micha{\l{}} Karpi{\'n}ski\authormark{1,*}}

\address{\authormark{1}Faculty of Physics, University of Warsaw, Pasteura 5, 02-093 Warszawa, Poland}

\email{\authormark{*}Michal.Karpinski@fuw.edu.pl} %% email address is required



\begin{abstract}
Spectral-temporal shaping of quantum light has important applications in quantum communications and photonic quantum information processing. Electro-optic temporal lenses have recently been recognized as a tool for noise-free, efficient spectral bandwidth manipulation of single-photon wavepackets. However, standard electro-optic time lenses based on single-tone modulation exhibit limited bandwidth manipulation due to material limitations on phase modulation amplitude. Here we numerically investigate the use of complex electro-optic temporal phase modulation patterns for bandwidth compression of light over multiple orders of magnitude and show the feasibility of their use in photonic interfaces for quantum network applications.
\end{abstract}

%\ocis{(270.0270) Quantum optics; (320.5520) Pulse compression; (260.2030) Dispersion.} % REPLACE WITH CORRECT OCIS CODES FOR YOUR ARTICLE, MINIMUM OF TWO; Avoid using the OCIS codes for “General” or “General science” whenever possible.
%For a complete list of OCIS codes, visit: https://www.osapublishing.org/oe/submit/ocis/

%%%%%%%%%%%%%%%%%%%%%%% References %%%%%%%%%%%%%%%%%%%%%%%%%

\bibliography{references}

\begin{thebibliography}{10}
\newcommand{\enquote}[1]{``#1''}

\bibitem{Kimble2008}
H.~J. Kimble, \enquote{The quantum internet,} {\protect\JournalTitle{Nature}}
  \textbf{453}, 1023--1030 (2008).

\bibitem{Reim2010}
K.~F. Reim, J.~Nunn, V.~O. Lorenz, B.~J. Sussman, K.~C. Lee, N.~K. Langford,
  D.~Jaksch, and I.~A. Walmsley, \enquote{{Towards high-speed optical quantum
  memories},} {\protect\JournalTitle{Nat. Photon.}} \textbf{4}, 218--221
  (2010).

\bibitem{Jensen2011}
K.~Jensen, W.~Wasilewski, H.~Krauter, T.~Fernholz, B.~M. Nielsen, M.~Owari,
  M.~B. Plenio, A.~Serafini, M.~M. Wolf, and E.~S. Polzik, \enquote{{Quantum
  memory for entangled continuous-variable states},}
  {\protect\JournalTitle{Nature Physics}} \textbf{7}, 13--16 (2011).

\bibitem{Lvovsky2009}
A.~I. Lvovsky, B.~C. Sanders, and W.~Tittel, \enquote{{Optical quantum
  memory},} {\protect\JournalTitle{Nat. Photon.}} \textbf{3}, 706--714 (2009).

\bibitem{Tiarks2016a}
D.~Tiarks, S.~Schmidt, G.~Rempe, and S.~D\"urr, \enquote{{Optical $\pi$ phase
  shift created with a single-photon pulse},} {\protect\JournalTitle{Science
  Advances}} \textbf{2}, e1600036 (2016).

\bibitem{Eckstein2011}
A.~Eckstein, B.~Brecht, and C.~Silberhorn, \enquote{{A quantum pulse gate based
  on spectrally engineered sum frequency generation},}
  {\protect\JournalTitle{Optics Express}} \textbf{19}, 13770--13778 (2011).

\bibitem{Campbell2014}
G.~T. Campbell, O.~Pinel, M.~Hosseini, T.~C. Ralph, B.~C. Buchler, and P.~K.
  Lam, \enquote{{Configurable unitary transformations and linear logic gates
  using quantum memories},} {\protect\JournalTitle{Physical Review Letters}}
  \textbf{113}, 063601 (2014).

\bibitem{Duan2001}
L.-M. Duan, M.~D. Lukin, J.~I. Cirac, and P.~Zoller, \enquote{{Long-distance
  quantum communication with atomic ensembles and linear optics},}
  {\protect\JournalTitle{Nature}} \textbf{414}, 413--418 (2001).

\bibitem{Ramelow2012}
S.~Ramelow, A.~Fedrizzi, A.~Poppe, N.~K. Langford, and A.~Zeilinger,
  \enquote{{Polarization-entanglement-conserving frequency conversion of
  photons},} {\protect\JournalTitle{Physical Review A}} \textbf{85}, 013845
  (2012).

\bibitem{McGuinness2010}
H.~J. McGuinness, M.~G. Raymer, C.~J. McKinstrie, and S.~Radic,
  \enquote{{Quantum frequency translation of single-photon states in a photonic
  crystal fiber},} {\protect\JournalTitle{Physical Review Letters}}
  \textbf{105}, 093604 (2010).

\bibitem{Rakher2011}
M.~T. Rakher, L.~Ma, M.~Davan{\c{c}}o, O.~Slattery, X.~Tang, and K.~Srinivasan,
  \enquote{{Simultaneous wavelength translation and amplitude modulation of
  single photons from a quantum dot},} {\protect\JournalTitle{Physical Review
  Letters}} \textbf{107}, 083602 (2011).

\bibitem{Zaske2012}
S.~Zaske, A.~Lenhard, C.~A. Ke{\ss}ler, J.~Kettler, C.~Hepp, C.~Arend,
  R.~Albrecht, W.-M. Schulz, M.~Jetter, P.~Michler, and C.~Becher,
  \enquote{{Visible-to-telecom quantum frequency conversion of light from a
  single quantum emitter},} {\protect\JournalTitle{Physical Review Letters}}
  \textbf{109}, 147404 (2012).

\bibitem{Wright2017}
L.~J. Wright, M.~Karpi{\'{n}}ski, C.~S{\"{o}}ller, and B.~J. Smith,
  \enquote{{Spectral shearing of quantum light pulses by electro-optic phase
  modulation},} {\protect\JournalTitle{Physical Review Letters}} \textbf{118},
  023601 (2017).

\bibitem{Clemmen2016}
S.~Clemmen, A.~Farsi, S.~Ramelow, and A.~L. Gaeta, \enquote{{Ramsey
  interference with single photons},} {\protect\JournalTitle{Physical Review
  Letters}} \textbf{117}, 223601 (2016).

\bibitem{Lavoie2013}
J.~Lavoie, J.~M. Donohue, L.~G. Wright, A.~Fedrizzi, and K.~J. Resch,
  \enquote{{Spectral compression of single photons},}
  {\protect\JournalTitle{Nat. Photon.}} \textbf{7}, 363--366 (2013).

\bibitem{Allgaier2017}
M.~Allgaier, V.~Ansari, L.~Sansoni, C.~Eigner, V.~Quiring, R.~Ricken,
  G.~Harder, B.~Brecht, and C.~Silberhorn, \enquote{{Highly efficient frequency
  conversion with bandwidth compression of quantum light},}
  {\protect\JournalTitle{Nature Communications}} \textbf{8}, 14288 (2017).

\bibitem{Karpinski2017}
M.~Karpi\'nski, M.~Jachura, L.~J. Wright, and B.~J. Smith, \enquote{{Bandwidth
  manipulation of quantum light by an electro-optic time lens},}
  {\protect\JournalTitle{Nat. Photon.}} \textbf{11}, 53--57 (2017).

\bibitem{Agha2014}
I.~Agha, S.~Ates, L.~Sapienza, and K.~Srinivasan, \enquote{{Spectral broadening
  and shaping of nanosecond pulses: toward shaping of single photons from
  quantum emitters},} {\protect\JournalTitle{Optics Letters}} \textbf{39},
  5677--5680 (2014).

\bibitem{Kues2017}
M.~Kues, C.~Reimer, P.~Roztocki, L.~R. {Cort{\'{e}}s}, S.~Sciara, B.~Wetzel,
  Y.~Zhang, A.~Cino, S.~T. Chu, B.~E. Little, D.~J. Moss, L.~Caspani,
  J.~Aza{\~{n}}a, and R.~Morandotti, \enquote{{On-chip generation of
  high-dimensional entangled quantum states and their coherent control},}
  {\protect\JournalTitle{Nature}} \textbf{546}, 622--626 (2017).

\bibitem{Fan2016}
L.~Fan, C.-L. Zou, M.~Poot, R.~Cheng, X.~Guo, X.~Han, and H.~X. Tang,
  \enquote{{Integrated optomechanical single-photon frequency shifter},}
  {\protect\JournalTitle{Nat. Photon.}} \textbf{10}, 766--770 (2016).

\bibitem{Mittal2017}
S.~Mittal, V.~V. Orre, A.~Restelli, R.~Salem, E.~A. Goldschmidt, and M.~Hafezi,
  \enquote{{Temporal and spectral manipulations of correlated photons using a
  time lens},} {\protect\JournalTitle{Physical Review A}} \textbf{96}, 043807
  (2017).

\bibitem{Lu2018}
H.-H. Lu, J.~M. Lukens, N.~A. Peters, O.~D. Odele, D.~E. Leaird, A.~M. Weiner,
  and P.~Lougovski, \enquote{{Electro-optic frequency beam splitters and
  tritters for high-fidelity photonic quantum information processing},}
  {\protect\JournalTitle{Physical Review Letters}} \textbf{120}, 030502 (2018).

\bibitem{Lu2018a}
H.-H. Lu, O.~D. Odele, D.~E. Leaird, and A.~M. Weiner, \enquote{{Arbitrary
  shaping of biphoton correlations using near-field frequency-to-time
  mapping},} {\protect\JournalTitle{Optics Letters}} \textbf{43}, 743--746
  (2018).

\bibitem{Kolner1994}
B.~H. Kolner, \enquote{{Space-time duality and the theory of temporal
  imaging},} {\protect\JournalTitle{IEEE Journal of Quantum Electronics}}
  \textbf{30}, 1951--1963 (1994).

\bibitem{Torres-Company2011}
V.~Torres-Company, J.~Lancis, and P.~Andr{\'{e}}s, \enquote{{Space-time
  analogies in optics},} {\protect\JournalTitle{Progress in Optics}}
  \textbf{56}, 1--80 (2011).

\bibitem{Salem2013}
R.~Salem, M.~A. Foster, and A.~L. Gaeta, \enquote{{Application of space-time
  duality to ultrahigh-speed optical signal processing},}
  {\protect\JournalTitle{Advances in Optics and Photonics}} \textbf{5},
  274--317 (2013).

\bibitem{Li2013}
B.~Li, M.~Li, S.~Lou, and J.~Aza\~{n}a, \enquote{{Linear optical pulse
  compression based on temporal zone plates},} {\protect\JournalTitle{Optics
  Express}} \textbf{21}, 16814--16830 (2013).

\end{thebibliography}
\bibliographystyle{osajnl}

%%%%%%%%%%%%%%%%%%%%%%%%%%  body  %%%%%%%%%%%%%%%%%%%%%%%%%% 

\section{Introduction}

Photons are excellent carriers of quantum information thanks to their weak interactions with the environment. In particular, photonic links can be used to interface distinct quantum systems into quantum networks \cite{Kimble2008}. However, quantum network nodes, such as quantum memories, quantum gates, or quantum light sources, may exhibit largely disparate spectro-temporal characteristics, including different emission or absorption wavelengths, as well as vastly different spectral linewidths and lineshapes. In particular, characteristic spectral bandwidths of quantum systems may differ by many orders of magnitude, from single MHz to hundreds of GHz \cite{Reim2010,Jensen2011,Lvovsky2009,Tiarks2016a,Eckstein2011,Campbell2014,Duan2001}. From quantum information point of view the most interesting are systems exhibiting single-photon nonlinearity, which lie in the single MHz bandwidth range, and also a vastly used sponantaneus parametric down-convertion sources of entangled photons with typical bandwidths of hundreds of GHz.
Only spectrally matched photons will efficiently interfere or will be absorbed by atoms in quantum memories or gates. Therefore one not only requires tools for quantum frequency conversion \cite{Ramelow2012,McGuinness2010,Rakher2011,Zaske2012,Wright2017,Clemmen2016}, but also for coherent, phase-only bandwidth manipulation of quantum light pulses, which does not involve lossy filtering. Single-photon bandwidth manipulation may be realized using nonlinear optical methods \cite{Lavoie2013,Allgaier2017}, or by means of electro-optic phase modulation \cite{Karpinski2017}. The all-fiber electro-optic approach enables deterministic bandwidth manipulation which is free from optical noise, making it especially well-suited for quantum light signals \cite{Agha2014, Karpinski2017, Wright2017, Kues2017, Fan2016, Mittal2017, Lu2018, Lu2018a}.
\begin{figure}[!t]
	\centering
	\includegraphics[height = 17em]{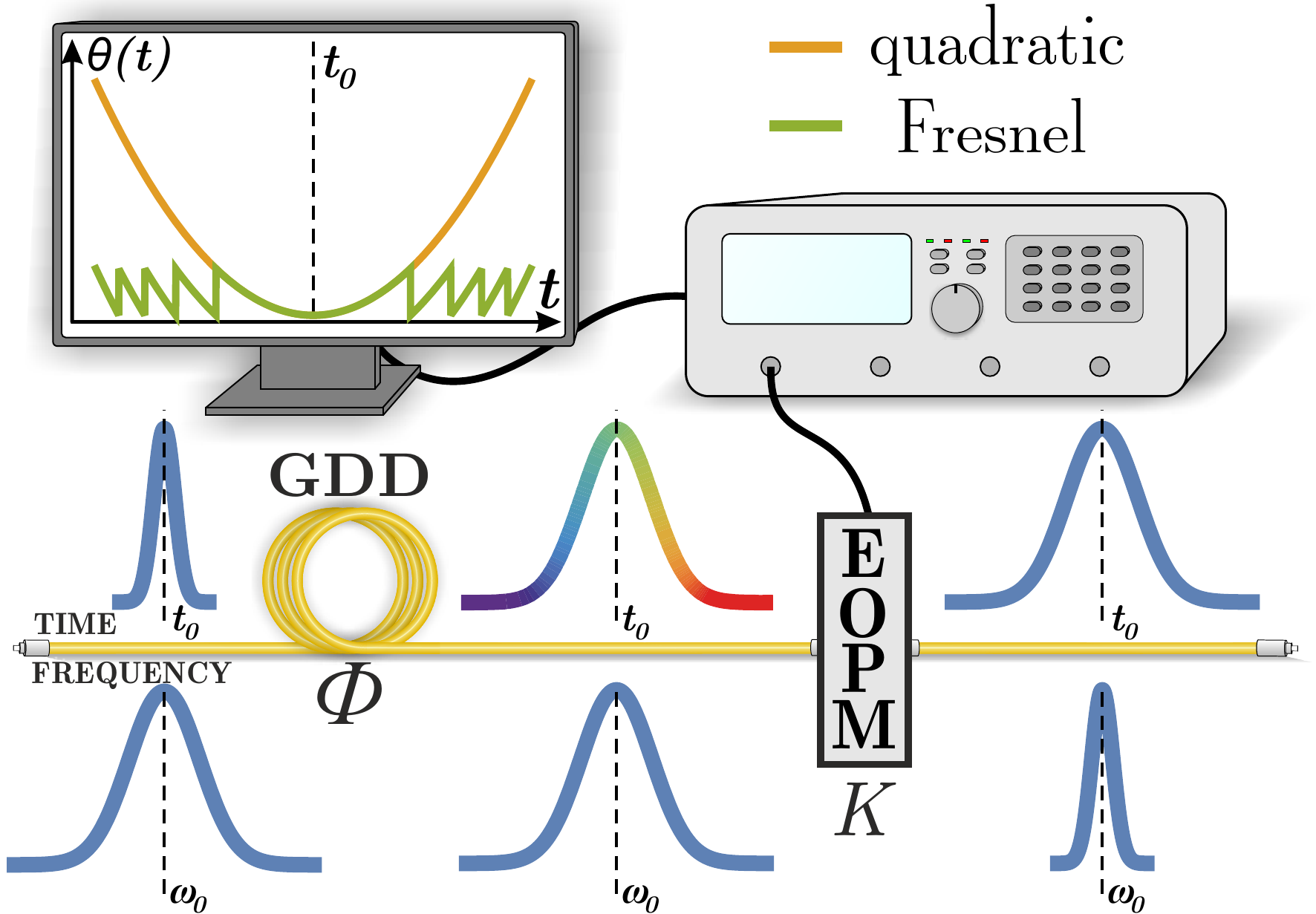}
	\caption{Principle of bandwidth conversion. A Gaussian optical pulse is chirped in second-order dispersive medium. Further it is subjected to quadratic temporal phase modulation (time lens) in an electro-optic phase modulator (EOPM) driven by an RF generator, resulting in compression of the pulse spectrum. The quadratic temporal phase profile may be replaced by a modulo-$2\pi$ profile, akin to the Fresnel lens principle. Time $t$ is given in a reference frame travelling at group velocity of central frequency of the pulse. }
	\label{fig:scheme}
\end{figure}

Experimental demonstrations of single-photon bandwidth compression to-date, both by electro-optic and nonlinear optical means, were only within the wideband, multi-GHz regime \cite{Lavoie2013,Karpinski2017,Allgaier2017}. An important necessity for the development of quantum technologies is to bridge narrowband, sub-GHz bandwidth pulses, compatible with trapped ions, or quantum gates based on Rydberg atoms, with multi-GHz bandwidths compatible with standard high bit-rate optical communication channels. Here we analyse the use of complex electro-optic temporal phase modulation patterns to perform deterministic bandwidth compression linking the wideband multi-GHz and narrowband multi-MHz bandwidth regimes. We show that using complex modulation patterns enables significantly increased bandwidth compression factors as compared with standard methods based on single-tone electro-optic phase modulation.

\section{Concept}

The principle of electro-optic bandwidth compression can be explained using the optical space-time (ST) duality \cite{Kolner1994,Torres-Company2011,Salem2013}, where paraxial propagation of light beam confined in space (diffraction) has its temporal counterpart in dispersive propagation in the time domain. The spatial spectrum of a Gaussian beam can be compressed by collimating it with a thin lens placed one focal length away from the beam waist. Upon diffractive propagation towards the lens the beam acquires a quadratic phase factor in transverse momentum; subsequently the spatial lens, in the paraxial approximation, imprints a quadratic phase in the transverse position variable, yielding a collimated output beam with an increased cross-section and narrower spatial spectrum. Compression of the spectral bandwidth of a Fourier-limited Gaussian optical pulse with spectral width $\sigma$, centred around $\omega_0$,  given by a spectral amplitude $\tilde{\psi}(\omega) = \pi^{-1/4} \sigma^{-1/2} \; \mathrm{exp}\left[ - \frac{(\omega - \omega_{0})^{2}}{2 \sigma^{2}} \right] $, can be realized as follows based on the ST analogy. Initially the pulse is sent through a length of dispersive medium, which chirps the pulse in time. Mathematically this corresponds to multiplication of the spectral amplitude by a quadratic spectral phase factor $\phi(\omega) = \Phi \left(\omega-\omega_{0}\right)^{2}\!/2$, where $\Phi$ is group delay dispersion (GDD). Further a time lens is applied to the pulse, in the form of quadratic temporal phase imprint $\theta(t) = Kt^{2}/2$, where $K$ is constant chirping factor, and $t$ is time in a reference frame travelling at the group velocity of the central frequency of the pulse. Spectral compression is achieved when the temporal analogue of the collimation condition is met: $K = 1/\Phi$. Bandwidth compression happens by the time lens introducing appropriate time-dependent spectral shifts onto the temporally dispersed spectral components of the pulse, as presented in Fig.\ \ref{fig:scheme}.For compression of Gaussian, Fourier-limited input pulse with width $\sigma_{\mathrm{in}}$ into target width $\sigma_{\mathrm{out}}$ a chirping factor $K = \sigma_{\mathrm{in}}\sigma_{\mathrm{out}}$ is needed \cite{Karpinski2017}.

The standard approach to realizing an electro-optic temporal lens involves aligning the optical pulse with the approximately quadratic region of sinusoidal temporal phase modulation \cite{Kolner1994,Torres-Company2011}. Here the achievable bandwidth compression factor is inherently limited by the achievable phase modulation amplitude: Let us assume a Fourier-limited, Gaussian pulse profile $\tilde{\psi}(\omega)$ and phase modulation profile of $\theta(t) = A \sin \left( 2 \pi f_{\mathrm{m}} t \right)$, where $A$ is the modulation amplitude and $f_{\mathrm{m}}$ is RF modulation frequency. The sinusiodal varying phase approximates a quadratic phase profile with chirping factor $K = 4 \pi^{2} f_{\mathrm{m}}^{2} A $ within a temporal aperture of duration $\alpha f_{\mathrm{m}}^{-1}$, where $\alpha$ is a constant describing the accuracy of the approximation \cite{Torres-Company2011}. The duration of a chirped pulse entering the electro-optic phase modulator (EOPM) is $\sigma_{\mathrm{in}}^{-1}\sqrt{1 + \sigma_{\mathrm{in}}^{4} \Phi^2} \approx \sigma_{\mathrm{in}} K^{-1}$ for large $\Phi = K^{-1}$ and should match the temporal aperture for the best performance of bandwidth compression, giving the input spectral width $\sigma_{\mathrm{in}} = K \alpha f_{\mathrm{m}}^{-1}$. The spectral width of the approximately Fourier-limited compressed pulse is proportional to the inverse of its duration, i.e. to the EOPM temporal aperture $\sigma_{\mathrm{out}}  = \alpha^{-1} f_{\mathrm{m}}$. It can be seen that the optimal compression factor $m = \frac{\sigma_{\mathrm{in}}}{\sigma_{\mathrm{out}}} = K \alpha^{2} f_{\mathrm{m}}^{-2} = 4 \pi^{2} \alpha^{2} A$ is proportional only to the phase modulation amplitude (with some constant factor), whose value is limited by material properties of electro-optic media \cite{Torres-Company2011}. In this approach the value of compression factor is on the order of 10, where losses from dispersive elements do not play a significant role.

One can increase the available temporal phase modulation aperture, while maintaining the value of $K$, by replacing the sinusoidal phase modulation by a quadratic modulo-$2\pi$ modulation profile increasing the available compression factor, as depicted in the inset of Fig. \ref{fig:scheme}, in analogy to spatial profiles of Fresnel lenses. This approach avoids phase modulation depth limitation, since only $2\pi$ modulation is needed. Here the main limitation, beside dispersion loss, is the electronic bandwidth {of the EOPM and driving electronics}  (throughout this work by electronic bandwidth we will mean the 3 dB bandwidth of electronic part of the setup, {which is given by the frequency, for which the frequency response of {electronics} drops by 3 dB}), which limits {the} temporal aperture of such {a} Fresnel lens.

{Let us analyze the effects of the limited electronic bandwidth of the setup. We assume an AWG, followed by an RF amplification circuit, is driving the EOPM. The slope of the }phase modulation profile $K t^{2}/2$, {is given by its derivative} $K t$, meeting at some point the electronic bandwidth $\Delta f$. Then the temporal aperture is {equal} to $ \Delta f/ K$ and {the} duration of the output optical pulse is given by {$\gamma \Delta f/K$}, where {$\gamma$} describes {the} overlap of electronic and optical signal. With good approximation the output spectral width is given by the inverse of duration of the pulse: {$\sigma_{\mathrm{out}} = K/ (\gamma \Delta f)$ yielding the compression factor: $m = \frac{\sigma_{\mathrm{in}}}{\sigma_{\mathrm{out}}} = \frac{K}{\sigma_{\mathrm{out}}^{2}} = \frac{\gamma^{2} \Delta f^{2}}{K}$}. While {$\gamma$} and $\Delta f$ {remain} physically limited parameters, the {chirping} factor $K$ can be chosen arbitrarily, leading to {an} unlimited compression factor in the ideal case. {This can be understood by noting that the maximum spectral shift required depends only on the input and is independent of the output bandwidth} However higher compression factors require higher amounts of dispersion with {higher} losses. In the ideal case, without losses, the spectral intensity of {the} output pulse {within the desired bandwidth $\sigma_{\mathrm{out}}$} is proportional to compression factor, however dispersion losses {reduce} it {exponentially} limiting the maximal {useful} compression factor.

\begin{figure}[!t]
	\centering
	\includegraphics[width = 0.79\linewidth]{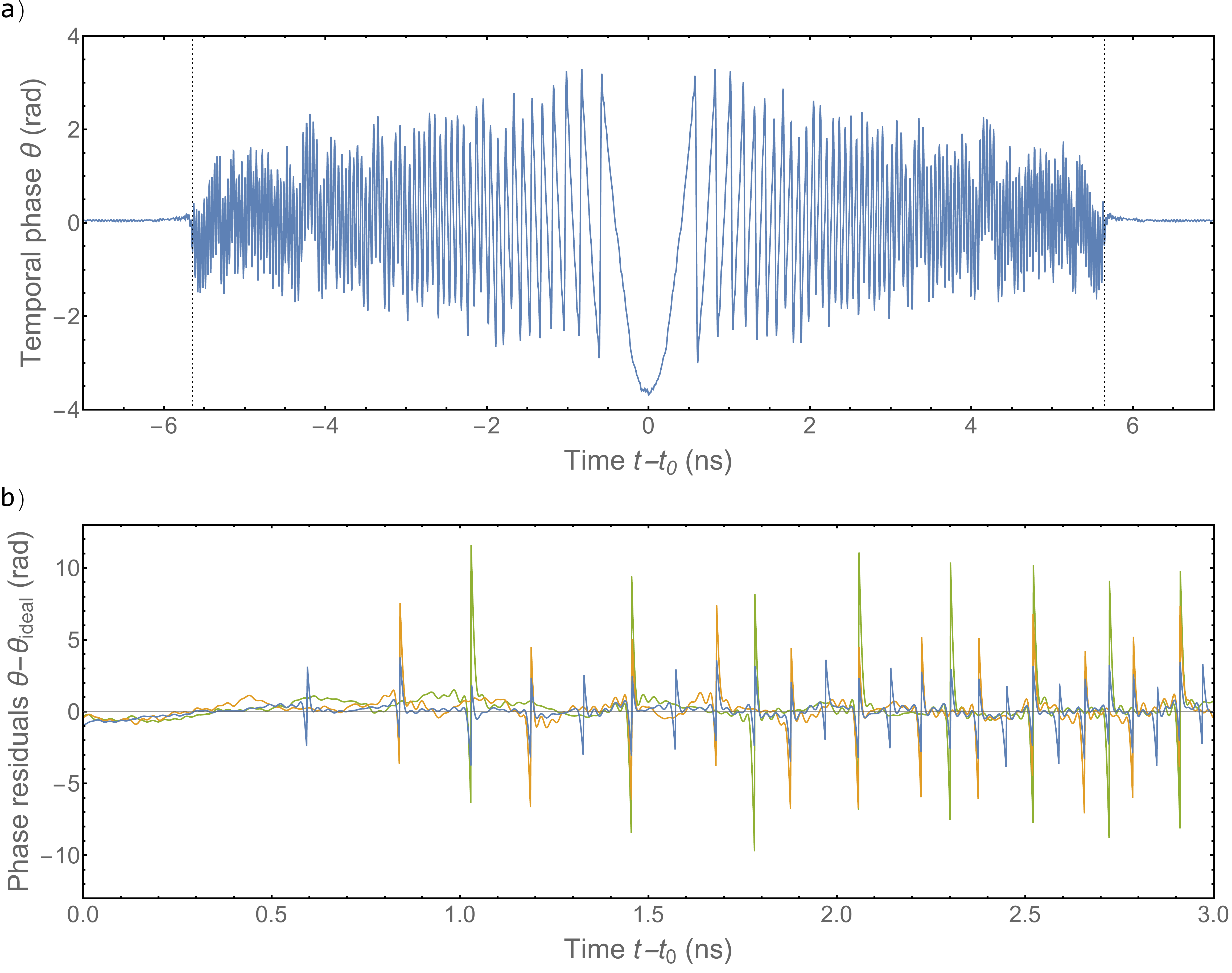}
	\caption{a) Realistic modulo-$2\pi$ Fresnel temporal phase profile {$\theta(t)$}	 for $K \approx 35.6 \, \mathrm{GHz}^{2}$ (corresponding to $\mathrm{GDD} \approx 28092 \, \mathrm{ps}^{2}$) including spectral responses of a 32-GHz electronic bandwidth arbitrary waveform generator (AWG), two RF amplifiers and EOPM (see text). b) Phase residuals for Fresnel orders of 1 ($2\pi$, blue), 2 ($4\pi$, orange) and 3 ($6\pi$, green). Please note that peaks of residuals for higher Fresnel orders are moved away from the center of the light pulse, while their height scales approximately with the Fresnel order. }
	\label{fig:order}
\end{figure}

\begin{figure}[!b]
	\centering
	\includegraphics[width = 0.78\linewidth]{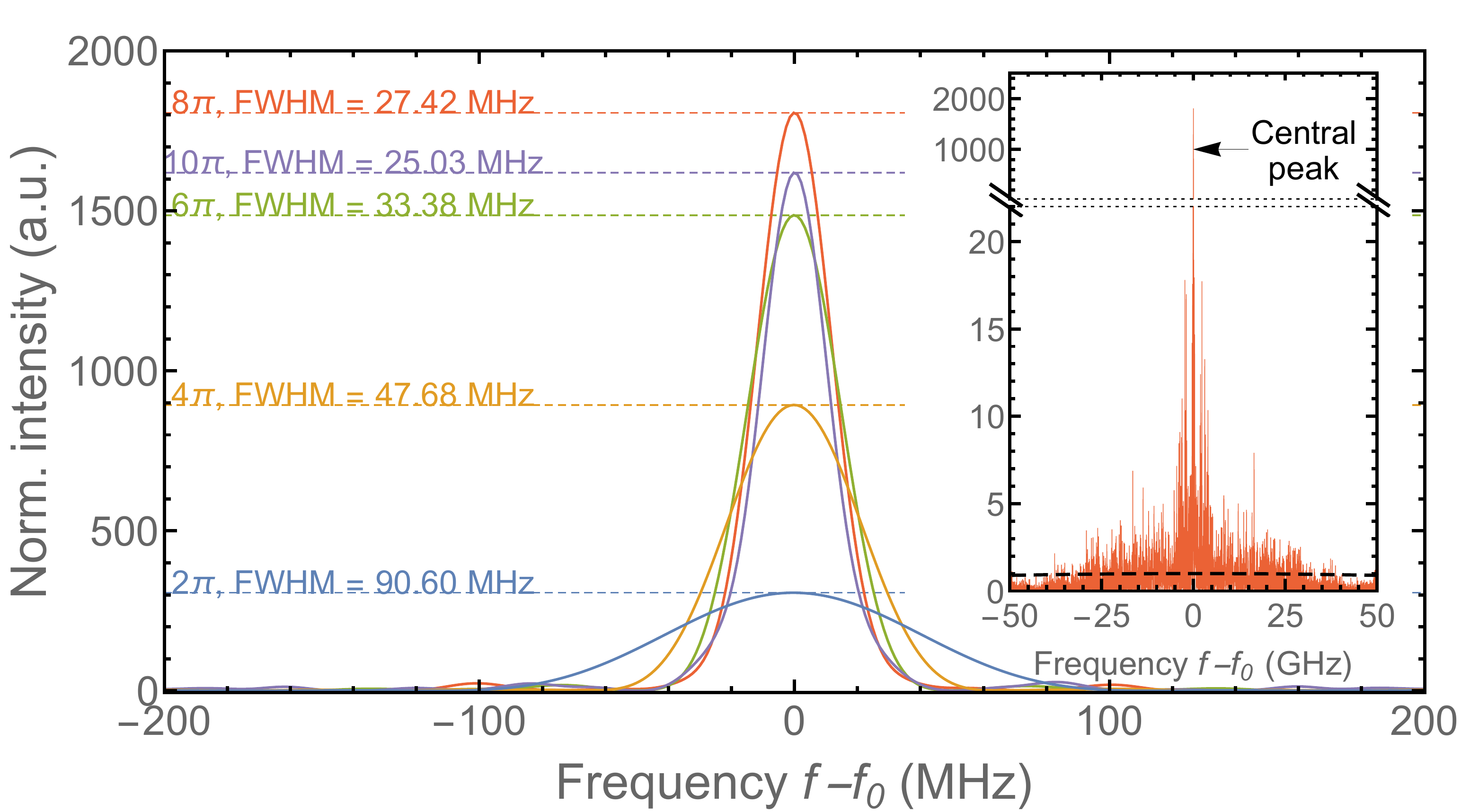}
	\caption{Spectral intensities of compressed pulses for different modulation depths (Fresnel orders). The input pulse full width at half maximum (FWHM) is $250$~GHz and the target FWHM is set to $10$~MHz. The output pulse is wider {than} the target spectral width due to non-{idealities} of the applied phase, shown in Fig \ref{fig:order}. Intensity is normalized to the input pulse. Inset: detail of the compressed spectrum for $8\pi$ modulation depth (solid, red) on top of the input pulse spectrum (dashed, black), normalized to the input pulse. Please notice different horizontal and vertical scales {in the inset}.  }
	\label{fig:signal}
\end{figure}

The challenge now lies in the need to generate a complex, very wide-band driving electrical signal{, that achieves the quadratic Fresnel-like lens depiced in the upper right inset in Fig.} \ref{fig:scheme}. Thus {the} sinusoidal RF generator needs to be replaced by an arbitrary waveform generator (AWG), and its non-ideal frequency response characteristics need to be {taken into account}. The temporal aperture in this approach is limited by the electronic bandwidth of the system. Experimental studies outside the context of quantum networks focused on short pulse generation from a continuous wave beam have shown the feasibility of spectral manipulation using complex temporal phase profiles in \cite{Li2013}. 

The quality of the time lens waveform is the first factor limiting the total performance of the bandwidth converter. The second perfomance limiting factor inherent to all approaches is loss from dispersive elements preceding the time lens. Here we first focus {on} the time lens quality only and then we introduce the losses from dispersive elements, which are inherent to any bandwidth compression realization based on a time lens.

\section{{Numerical simulations}}

We perform numerical simulations of an electro-optic bandwidth converter based on a Fresnel temporal lens. We assume a Gaussian pulse with an initial bandwidth of $\sigma_{\mathrm{in}}$. The pulse is chirped to the desired output temporal duration $\sigma_{\mathrm{out}}^{-1}$, by multiplying its spectral amplitude by a quadratic spectral phase in the frequency domain. We simulate realistic generation of EOPM driving signal $V(t)$ in order to obtain a realistic temporal phase modulation profile $\theta(t)$. To this end we include the following distortions: temporal and amplitude discretization, frequency response (electronic bandwidth) of the AWG, amplifiers and EOPM, noise and possible saturation of the amplifier. Further, the duration of a phase waveform, corresonding to the temporal aperture of the time lens, is limited by the maximal achievable signal slope. In our simulation we generate the input signal in frequency domain by discretizing it into $N = 2^{24}$ points. We use fast Fourier transforms (FFT), implemented on a graphics processing unit (GPU) using CUDA, to convert between time and frequency domains in order to apply appropriate temporal and spectral phases. {The discretization results in $ 0.5 \; \mathrm{ps}$ resolution in the temporal domain and $\sim \! 0.06 \; \mathrm{MHz}$ resolution in the frequency domain.}

The procedure to generate the phase modulation signal consists of programming the AWG, subsequent generation of the signal by the AWG and its further amplification. The signal is distorted at each step, starting from the need to digitize the desired waveform at the programming stage, through generation noise and frequency response of the AWG, amplifiers and EOPM. In particular our model includes discretization of the signal in time (according to the AWG sampling rate) and amplitude (according to the effective number of bits, ENOB, of the AWG, which also captures noise of the AWG), applying a temporal aperture (given by the electronic bandwidth of the AWG), and applying the frequency responses of the amplifiers and EOPM. {We also modify (precompensate) the generated waveform at the programming stage to counter the effects of non-flat frequency response of the elements.} 

\begin{figure}[!b]
	\centering
	\includegraphics[width=0.75\linewidth]{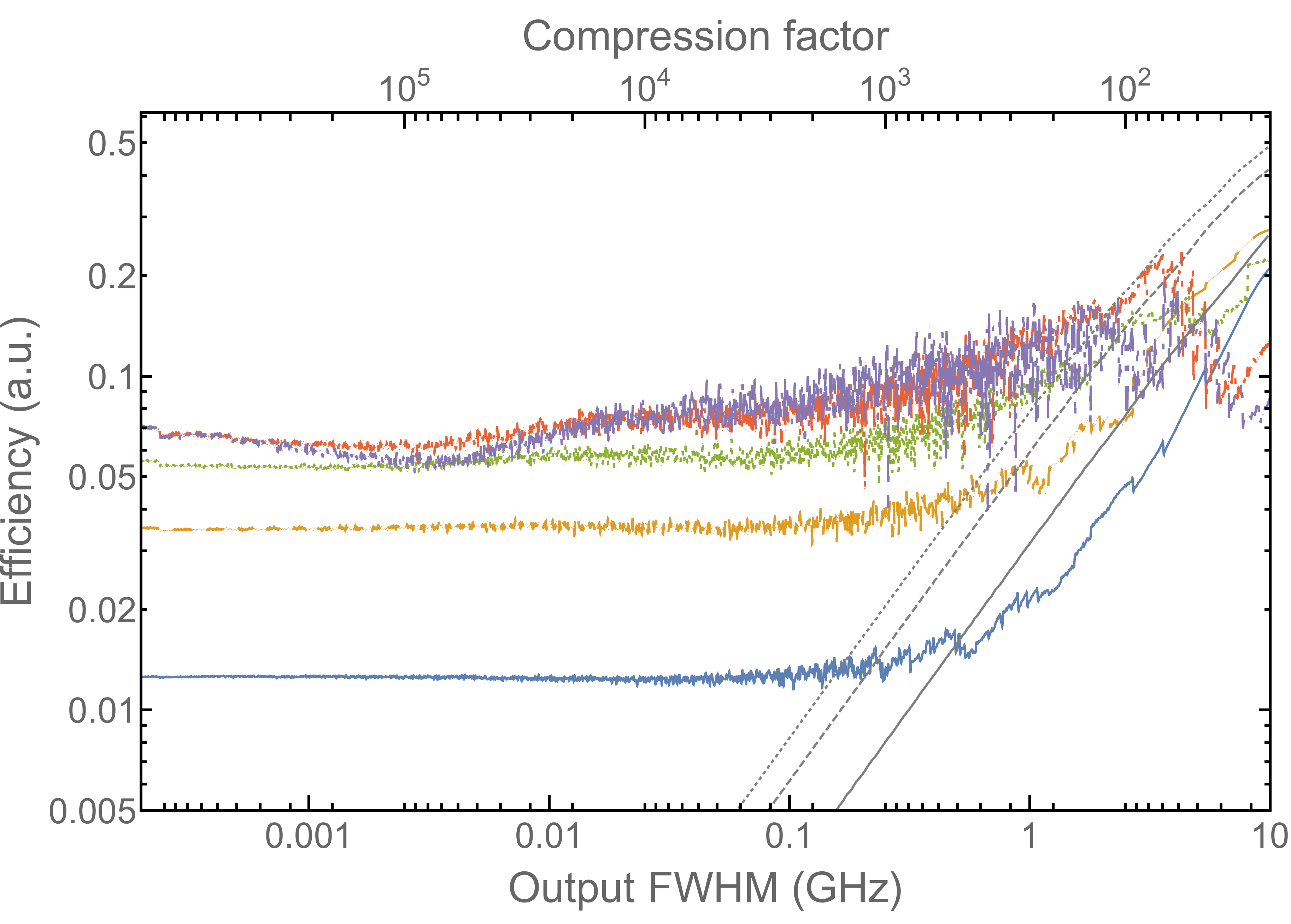}
	\caption{Efficiency of bandwidth compression for FWHM {spectral} intensity bandwidth of $250$~GHz, for Fresnel orders of 1 ($2\pi$, blue, solid), 2 ($4\pi$, orange, dashed),  3 ($6\pi$, green, dotted), 4 ($8\pi$, red, dot-dashed) and 5 ($10\pi$, violet, double dot-dashed). {For comparision a time lens using single-tone modulation with amplitude of 10 rad (solid, grey), 25 rad (dashed grey) and 40 rad (dotted, gray) is shown.  } }
	\label{fig:eff}
\end{figure}

{More specifically the procedure is as follows. First we calculate the ideal Fresnel waveform in the temporal domain. Then we use FFT to obtain its spectrum, which is precompensated by multiplying it by the inverse of the combined frequency response of the setup within the bandwidth of the AWG. Afterwards we use the inverse FFT (IFFT) to obtain the temporal waveform again. Next we discretize the waveform using $\lfloor 2^{\mathrm{ENOB}}\rfloor$ vertical steps with duration given by the inverse of the sampling rate of the AWG. We find maximal and minimal values of the waveform which give the current waveform amplitude. The possible values of these steps are given by evenly distributing $\lfloor 2^{\mathrm{ENOB}}\rfloor$ values within the waveform amplitude. Then we sample the waveform by setting values of the samples to the vertical steps closest to the values of the waveform before discretization, in the center of the sample. This sampling is done from the center of the waveform $(t=0)$ to the outside $(t\neq0)$, until it reaches the rise/fall time of the AWG, which is given by its bandwidth. After this point we set the waveform to 0, which introduces the temporal aperture. Then we apply the frequency response, by first using FFT to obtain the spectrum of generated waveform after sampling. Subsequentally the spectrum is multiplied by the frequency response of the AWG, preampilifer, amplifier and EOPM, where frequency response of amplifiers also introduces gain. The final phase waveform is obtained by using IFFT. Due to non-flat frequency characterisitics of the electronic circuit, the amplitude of the waveform has to be optimized by rescaling it to reach the best performance of the time lens. Therefore we obtain the correct chirping factor $K$.}
	
{Electronic noise in the system consisting of a series of high-gain amplifiers is dominated by the noise of the first element of the series as per formula ??. In our case the key source of RF amplitude noise is thus the AWG. The noise of the AWG is included through the reduced AWG ENOB. We assume perfect temporal synchronization of the setup given that RF signals can be synchronized to optical signals with timing jitters orders of magnitued lower then the AWG tempral resulution. [Azana?]}

We use the parameters of commercially available elements:{ the Keysight M8196A AWG with $32$~GHz electronic bandwidth, sampling rate of $92$~GS/s and ENOB of 5.5 (45 amplitude steps). The AWG is followed by the Keysight N4985A-S50 preamplifier and the RF-LAMBDA RFLUPA01G31GB power amplifier with electronic bandwidth of range 0.2-35 GHz and saturation power of 32.5~dBm, which is followed by the EO-SPACE PM-5SE-10-PFU-PFU-UV waveguided EOPM with half-wave voltages of $3\,\mathrm{V}$ at $1\,\mathrm{GHz}$  and $6.35\,\mathrm{V}$ at $32\,\mathrm{GHz}$.}

\begin{figure}[!b]
	\centering
	\includegraphics[width=0.75\linewidth]{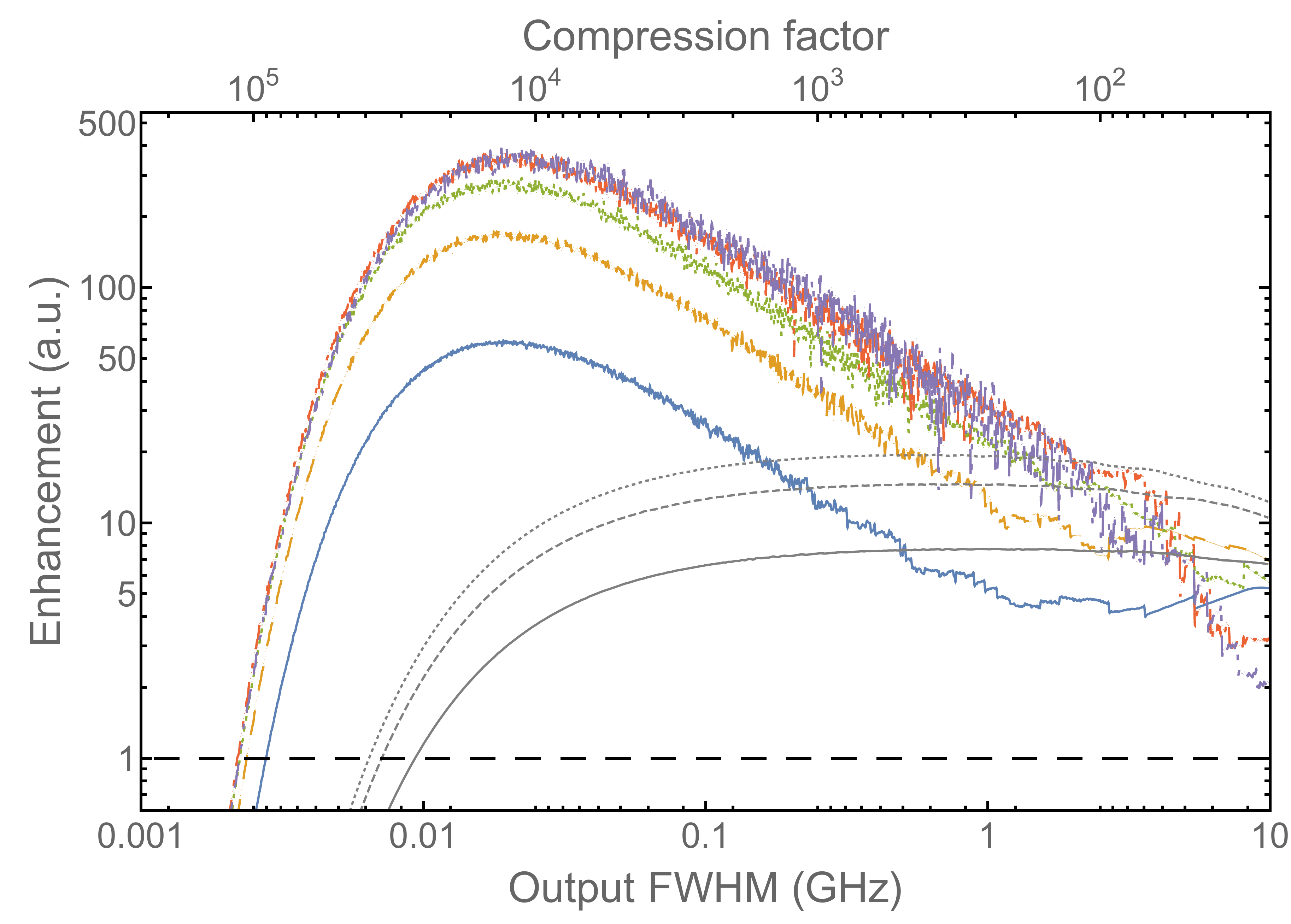}
	\caption{ Enhancement for a $250$~GHz FWHM spectral intensity bandwidth input pulse. The lines correspond to Fresnel order of 1 ($2\pi$, blue, solid), 2 ($4\pi$, orange, dashed), 3 ($6\pi$, green, dotted), 4 ($8\pi$, red dot-dashed) and 5 ($10\pi$, violet, double dot-dashed). Losses associated with dispersive propagation of $3$~dB per 10000 $\mathrm{ps}^{2}$ were assumed. }
	\label{fig:enh}
\end{figure}

The bandwidth converter performance can be further increased by using modulation depths higher than $2\pi$: $4\pi$, $6\pi$ or in general $2 k \pi$, where $k \in \mathbb{N}$ is the Fresnel order. It can be achieved by generating a $Kt^{2}/2k$ modulo-$2\pi$ waveform in the AWG stage and subsecuently amplifying it $k$ times more in the amplifier stage, obtaining $2\pi k$ modulation and chirping factor $K$. It results in the phase waveform scaled $k$ times both in time and amplitude yielding larger temporal apertures, which increases the compression factor.  Another important aspect are wrapping points -- points of phase jumps from $k\pi$ to $-k\pi$, which create distortions of the signal due to a limited electronic bandwidth of the generator. Higher Fresnel orders give $k$ times fewer wrapping points, but with $k$-times higher distortion of the phase. Also, since the modulation depth is $k$ times higher, while the number of amplitude resolution steps is constant, the  vertical phase resolution is smaller. Further achieving high Fresnel orders may be prevented by saturating the wideband RF amplifier or reaching maximal acceptable voltage for the EOPM. We use numerical simulations to find the optimal balance of these effects.

\section{Results}	

In Fig.\ \ref{fig:order}(a) we present a precompensated temporal phase waveform for a time lens with $K \approx 35.6 \, \mathrm{GHz}^{2}$, modulo $2\pi$,  which corresponds to compression from 250 GHz to 10 MHz in the ideal case. In Fig.\ \ref{fig:signal} we present the spectral intensity of a $250$~GHz bandwidth optical pulse (inset -- black, dashed) whose bandwidth is compressed by chirping it in an ideal second order dispersive medium with $\Phi = 1/K$, and subjecting {it} to phase modulation with the above waveform. The  spectrum of the output signal  (blue, solid) is shown in {Fig.} \ref{fig:signal}. The obtained output FWHM is $90.60$~MHz, which is lower than the set FWHM due to limited temporal {aperture} and imperfections of the generated phase. One can also see a remaining, uncompressed signal on both sides of the central peak{, which has not been fully shifted towards the central peak due to nonidealities of the RF modulation} -- in realistic applications it can be filtered out, see inset of Fig. \ref{fig:signal}. 

\begin{figure}[!b]
	\centering
	\includegraphics[width=0.7\linewidth]{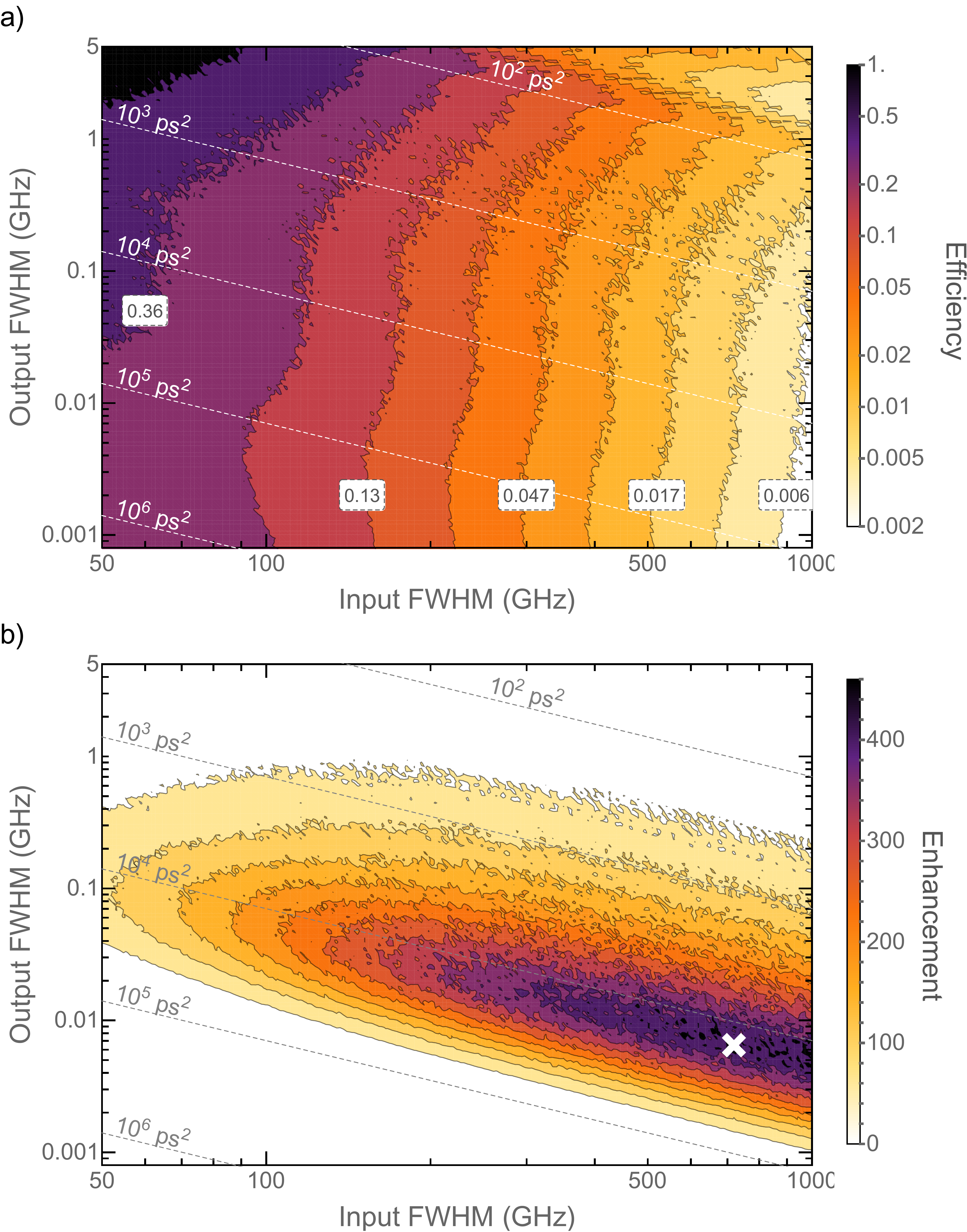}
	\caption{ Dependency of a) efficiency and b) enhancement of the bandwidth converter with the modulation depth of $8\pi$ on input and output FWHM spectral intensity bandwidths. In (b) the white cross marks the maximal enhancement in given input and output widths range with value of 449 for compression from 719~GHz to 7.8~MHz. The dashed lines show constant amount of needed dispersion $\Phi$.  }
	\label{fig:maps}
\end{figure}

In order to quantify the bandwidth converter performance we introduce two measures. We consider an input pulse subjected to the bandwidth
converter employing a lossless dispersive medium. It is followed by a rectangular spectral filter of a given spectral width. The chirping factor of the bandwidth converter is set such, that the full width at half-maximum (FWHM) spectral intensity of the output pulse is matched to the width of the filter for an ideal quadratic phase modulation profile. We define the efficiency as the ratio of the total intensity after the spectral filter for the realistic phase profile to the intensity in the case of the ideal quadratic profile. The efficiency captures the impact of phase waveform nonidealities and temporal aperture on the compression process. The second measure, the enhancement, compares a scheme where a bandwidth converter is used to the case where there is no spectrum compression. It is a ratio of total intensities after the rectangular spectral filter for both the cases. Enhancement also captures losses originating from dispersive elements, providing the total system performance of the bandwidth converter.

In Fig.\ 4 we plot the efficiency for input FWHM spectral intensity bandwidth of $250$~GHz and a range of output bandwidths and Fresnel orders. The efficiency is initially fast decreasing with an increasing compression factor. This is because in this regime only a small amount of {group delay dispersion $\Phi$ is used}. Thus the slopes in the Fresnel phase waveform $\phi(t) = K t^{2}/2 = t^{2}/2\Phi$ are very steep, leading to a very small aperture of the time lens. For higher compression factors, where more dispersion is needed, the efficiency is reaching a constant value. The improvement of using higher Fresnel order is approximately $k$-fold up to $k=3$.

While the efficiency quantifies the performance of the time lens alone, the enhancement takes into account also losses arising from the dispersive element forming the bandwidth converter.{ Therefore we define the quantity $\Theta$, which describes amount of the group delay dispersion, which introduces $3$~dB loss and which value depends on used dispersive elements. Then the transmission for any value of $\Phi$ is given by $T = 0.5^{\Phi/\Theta}$. Here we envision using multiple commercially available units of chirped fiber Bragg gratings (CFBG), which exhibit $\Theta$ up to 10000 $\mathrm{ps}^{2}$ based on the specifications of commercially available products}	. The plot in Figure \ref{fig:enh} shows a linear scaling of enhancement in terms of compression factor, up to some maximal value dependent on assumed $\Theta$. In the regime of high compression factor the time lens efficiency is constant, as shown in Figure \ref{fig:eff}, but losses from dispersive elements are growing exponentially with compression factor. Therefore it is important to use low-loss, high-dispersive elements, such as chirped fiber Bragg gratings. The improvement of $\Theta$ will increase performance of our bandwidth converter for given pair of input-output widths. It will also allow to achieve lower target widths of optical pulses. We analyzed a range of 50--1000 GHz of input FWHMs and a range of 0.1 MHz -- 5 GHz of output FWHMs, see Figure \ref{fig:maps}, {where we show (a) the efficiency of the time lens and (b) the enhancement of using bandwidth converter. In Fig. }\ref{fig:maps}{(a) one can notice the decrease of the efficiency along lines of the constant dispersion $\Phi$. Along these lines the same phase waveform is generated with chirping factor $K = 1/\Phi$. However the waveform is used for different pairs of input-output widths of optical pulses, therefore the phase waveform covers smaller part of the pulse, which causes the decrease of efficiency. The enhancement adds also losses from dispersive elements which limits the enhanement in the low-input low-output width region}. We show that {the} use of complex (Fresnel) waveforms allows a 2 orders of magnitude gain in single-photon flux when interfacing bandwidths differing by more than 3 orders of magnitude.

\section{{Conclusions}}

In conclusion we show that temporal lenses based on complex electro-optic modulation patterns enable spectral bandwidth conversion over multiple orders of magnitude, linking wideband GHz bandwidth pulses, compatible with optical fiber communication channels, with narrowband MHz emitters and absorbers, such as trapped ions or Rydberg atomic media. We use numerical methods to analyse the effects of non-ideal generation of phase modulation and {indicate} the feasibility of experimental implementation of the method with commercially available RF components. The presented approach permits {an} all fiber implementation, is easily reconfigurable and wavelength-tunable and uses only determinstic, linear operations, making it naturally suited for single-photon pulses.

\section*{Acknowledgments}
We thank  C. Becher, A.~O.~C.\ Davis, B.~J.\ Smith, and V.\ Thiel for insightful discussions. 

\section*{Funding}
This research was funded inpart by the National Science Centre of Poland (project No. 2014/15/D/ST2/02385) and inpart by the HOMING programme of the Foundation for Polish Science (project no. Homing/2016-1/4), co-financed by the European Union under the European Regional Development Fund.

\end{document}